\documentclass{optica-article}

\journal{opticajournal} 

\articletype{Research Article}

\usepackage{lineno}
\usepackage[separate-uncertainty=true,multi-part-units=single]{siunitx}

\newcommand{\abs}[1]{\left| #1 \right|}

\begin{document}

\title{Demonstration of a tandem lens for producing shaped laser-ionized plasmas for plasma wakefield acceleration}

\author{R. Ariniello,\authormark{1,2,*} V. Lee,\authormark{2} and M. D. Litos\authormark{2}}

\address{\authormark{1}SLAC National Accelerator Laboratory, Menlo Park, California 94025, USA\\
\authormark{2}Center for Integrated Plasma Studies, Department of Physics, University of Colorado Boulder, Boulder, Colorado 80309, USA\\}

\email{\authormark{*}robert.ariniello@colorado.edu}


\begin{abstract*} 
We demonstrate a tandem lens optical setup, comprising two diffractive optics, that focuses a high-power ultrafast laser with a shaped on-axis intensity profile, producing a meter-long Bessel focus. The intended use of the optical setup is to produce a laser-ionized plasma source for plasma wakefield acceleration. By controlling the on-axis intensity, the density profile of the plasma ramps at the entrance and exit of the plasma can be tailored to optimize matching of the electron beam into the plasma. In addition to demonstrating the optical system, we describe the algorithm used to calculate the lens phases and present detailed calculations of the lenses' expected performance. 
\end{abstract*}

\section{\label{sec:intro}Introduction}

Plasma wakefield accelerators (PWFA) can generate accelerating gradients three orders of magnitude larger than conventional accelerators \cite{hogan2005multi,blumenfeld2007energy}, making them a promising candidate for an energy frontier collider to support high-energy particle physics \cite{adli2013beam,foster2023hybrid,asai2024exploring} or as a compact, low-cost alternative to existing light sources \cite{nakajima2008towards, wang2021free, pompili2022free, labat2023seeded}. In a beam driven PWFA, an ultrarelativistic ``drive" beam of charged particles excites a density wave in a plasma, producing strong transverse and longitudinal electromagnetic wakefields that accelerate and focus a trailing "witness" beam. 

A PWFA with multi-GeV electron drivers requires plasmas with lengths from \qty{10}{cm} up to \qty{4}{m}, widths from \qty{100}{\micro\meter} to \qty{200}{\micro\meter}, and plasma densities from \qty{5e15}{cm^{-3}} to \qty{1e17}{cm^{-3}}. To reach these parameters, experiments have primarily used three plasma sources: First, lithium vapor ovens \cite{muggli1999photo} have demonstrated high-gradient \cite{hogan2005multi, blumenfeld2007energy} and high drive-to-witness efficiency with low energy spread \cite{litos2014high, litos20169} as well as ionization injection of bright beams \cite{oz2007ionization, vafaei2016evidence}. However, due to overheating, they are limited to repetition rates $\leq\qty{1}{Hz}$ when operating at high energy transfer efficiency. Second, argon discharges have enabled energy spread preservation \cite{lindstrom2021energy} and emittance preservation \cite{lindstrom2024emittance} at low, $<5\%$, energy gain of the witness, along with high drive-to-wake energy transfer efficiency \cite{pena2024energy} with a short, high-density plasma, but face challenges in repetition rate and producing uniform, long plasmas at densities suitable for energy doubling the witness. Finally, laser-ionized hydrogen plasmas have been used to produce high-brightness beams with the plasma photocathode technique \cite{deng2019generation}, but have not yet been successfully used with an externally injected witness beam due to challenges with plasma width and alignment. None of the plasma sources allow easy control over the ramp shapes and plasma length, both of which are crucial for high-quality acceleration \cite{xu2016physics, litos2019beam, ariniello2019transverse, zhao2020emittance, ariniello2022chromatic}. If the alignment and density ramp tunability problems can be overcome, laser-ionized plasma sources have the advantages of full optical access to the plasma (useful for diagnostic purposes \cite{zgadzaj2020dissipation, gilljohann2019direct, lee2024temporal}), a kHz potential repetition rate, and immunity to electron beam damage.

The plasma ramps and plasma length can be controlled and tuned by shaping the laser intensity on-axis. Near arbitrary on-axis intensity profiles can be produced using a superposition of Bessel beams \cite{vcivzmar2009tunable}. This approach has been demonstrated using a spatial light modulator (SLM) combined with a pinhole for a CW laser \cite{vcivzmar2009tunable, lu2022tunable, fontaine2019attenuation, yan2021non} and an ultrashort pulsed laser \cite{ouadghiri2016arbitrary}, but the approach is not practical for the several TW-scale laser pulses required to ionize a meter-scale plasma. For one, available SLMs do not have sufficient size and damage threshold for such a high-power laser, and two, the energy efficiency of the technique is $<10\%$. Another shaping technique has been proposed that utilizes a pair of optics, with the shaping based on a ray optics approach \cite{honkanen1998tandem, dharmavarapu2018diffractive}. With this technique, the optical setup could be adapted to high power lasers, but the ray optics approach results in diffraction ringing and fidelity limitations that make the shaping unsuitable for PWFA applications. Further, for the meter-scale depth of focus and pulse durations relevant for PWFA plasma sources, chromatic effects in the ultrashort pulse become important.

To overcome these challenges, we combine the tandem lens of Honkanen and Turunen \cite{honkanen1998tandem} with the fully diffractive calculation of Cizmar and Dholakia \cite{vcivzmar2009tunable} to create our novel optical setup. To do this, we developed an algorithm to calculate the phases of both lenses from the diffractive calculation and extended the diffractive treatment to ultrafast pulses. We verified the lenses' performance by comparing the measured results to predictions from a combination of a ray optics efficiency calculation and a physical optics simulation. Using our optical setup, we demonstrate the laser ionization of an \qty{1.35}{Torr} $\mathrm{H_2}$ gas. Our results demonstrate that a laser-ionized plasma source has significant potential as a plasma source for a PWFA.

\section{\label{sec:lensDesign}Lens Design Algorithm}

\begin{figure}[htbp]
  \centering
  \includegraphics[width=5.25in]{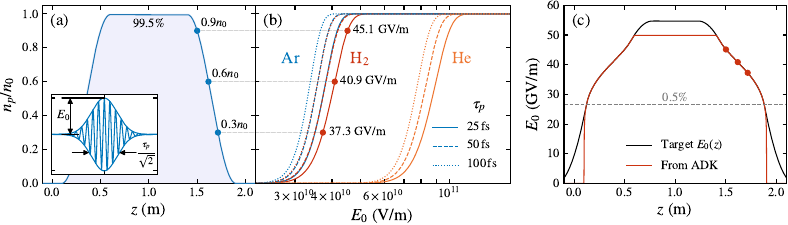}
  \caption{Calculation of the electric field required to create a longitudinally shaped plasma. (a) Target on-axis plasma density along with the pulse shape of the ionizing laser (inset), here a \qty{0.8}{m} long plasma with \qty{0.5}{m} cosine ramps and an $\qty{800}{nm}$ Gaussian pulse with FWHM duration $\tau_p$. (b) Ionization fraction for $\mathrm{Ar}$, $\mathrm{H_2}$, and $\mathrm{He}$ as a function of the peak electric field $E_0$, from Eq.~(\ref{eq:ionizationFrac2}). The on-axis electric field required to achieve a given ionization fraction can be found by numerically inverting the appropriate curve; three representative points are displayed for $\mathrm{H_2}$. (c) Required electric field strength along the optical axis. The target field must be smooth, accomplished by increasing the field strength in the fully-ionized region and tapering the field to zero below 0.5\% ionization (black). \label{fig:IonizationFraction}}
\end{figure}

The lens design task is to create an optical setup that will produce a laser-ionized plasma with a prescribed on-axis plasma density $n_p(z)$ from an incoming laser pulse with an electric field given, in the scalar approximation, by $E_{in}(r, t) = \mathcal{N}E_\perp(r) E_t(t)e^{-i\omega_0 t}$, where $\mathcal{N}$ is an overall normalization factor, $E_\perp(r)$ is the transverse profile of the pulse, $E_t(t)$ is the temporal envelope, and $\omega_0$ is the angular frequency of the laser. Here, $E_\perp(r)$ and $E_t(t)$ are unit-less. The first step is to calculate the amplitude of the on-axis electric field $E_0(z)$ needed to ionize a plasma with density $n_p(z)$ from a neutral gas with density $n_0$. The ionization fraction after the laser has passed is given by~\cite{lai2017experimental}
\begin{equation} \label{eq:ionizationFrac}
    \frac{n_p}{n_0} = 1-e^{-\int w(t)\,dt},
\end{equation}
where $w(t)$ is the time-dependent ionization rate. Assuming that the pulse envelope remains unchanged as the pulse propagates, then $w(t)$ can be calculated using the ADK model \cite{Ammosov1986}. Inserting into Eq.~(\ref{eq:ionizationFrac}) we get
\begin{equation} \label{eq:ionizationFrac2}
    \frac{n_p}{n_0} = 1-\exp\left(-\int w_{\!A\!D\!K}\left[E_0E_t(t)e^{-i\omega_0 t}\right]\,dt\right),
\end{equation}
where the functional form of $w_{\!A\!D\!K}$ is given in App.~\ref{app:ADK}. Equation~(\ref{eq:ionizationFrac2}) can be integrated numerically for different values of $E_0$, providing $n_p/n_0$ as a function of $E_0$. Interpolation can then be used to invert the function, yielding $E_0(z)$ from $n_p(z)$. An example of this process is shown in Fig.~\ref{fig:IonizationFraction} for a Gaussian pulse ionizing $\mathrm{Ar}$, $\mathrm{H_2}$, and $\mathrm{He}$. The target ionization fraction must be less than one because Eq.~(\ref{eq:ionizationFrac2}) only asymptotically approaches complete ionization; 99.5\% works well.

Note that the ADK model applies in the tunneling regime, where the Keldysh parameter $\gamma=\omega_0/\omega_t\ll 1$, here $\omega_t=\frac{eE_0}{\sqrt{2m_e\xi_i}}$, $e$ is the elementary charge, $m_e$ is the electron mass, and $\xi_i$ is the ionization energy of the outer electron in the neutral gas \cite{Ammosov1986}. Outside of the tunneling regime, the PPT theory may be used instead; see Ref.~\cite{perelomov1966ionization}. 

With $E_0(z)$ determined, the second step is to calculate the electric field on the $z=0$ plane that, when propagated forward in $z$, will produce the desired field amplitude $E_0(z)$ along the optical axis. We extend the approach of Ref.~\cite{vcivzmar2009tunable} to include the bandwidth of an ultrashort pulse. Since the incoming laser pulse is assumed to be cylindrically symmetric, we begin with the general solution to the scalar wave equation in cylindrical coordinates, which is a superposition of Bessel beams,
\begin{equation} \label{eq:besselBeam}
    E(r, \phi, z, t) = \sum_m\int_{-\infty}^\infty d\omega\int_0^\infty k_r\,dk_r\,S_m(k_r, \omega) J_m(k_rr)e^{i(k_zz+m\phi-\omega t)},
\end{equation}
where $m$ is the azimuthal mode number, $J_m$ are the Bessel functions of the first kind, and $k_r$ and $k_z$ are the radial and longitudinal projections of the wavevector, respectively. They are related to the wavenumber $k=\omega n_g/c$, with $n_g$ being the index of refraction of the gas and $c$ the speed of light in vacuum, by 
\begin{equation} \label{eq:krkz}
    k^2=k_r^2+k_z^2.
\end{equation}
The quantity $S_m(k_r, \omega)$, known as the spectrum of spatial frequencies, is the amplitude of each Bessel beam.

Only the $m=0$ Bessel function is of interest, as $E_0(z)$ is nonzero along the $z$ axis. We can solve for $S_0(k_r, \omega)$ in terms of $E_0(z)$ by setting $r=0$ in Eq.~(\ref{eq:besselBeam}) and changing the integration variable to $k_z$ using $k_z=\sqrt{k^2-k_r^2}$. The result is
\begin{equation} \label{eq:onAxisField}
    E(0, z, t) = \int_{-\infty}^\infty d\omega\int_0^{k} k_z\,dk_z\,S'_0(k_z, \omega)e^{i(k_zz-\omega t)},
\end{equation}
where $S'_0(k_z, \omega)=S_0\big(\sqrt{k^2-k_z^2}, \omega\big)$ and we have truncated the integration range to $0$ to $k$, eliminating backwards propagating and evanescent modes. Because the on-axis field is the inverse Fourier transform of $S'_0(k_z, \omega)$, the spectrum of spatial frequencies can be found by Fourier transforming the target on-axis field and performing the $k_z$ to $k_r$ coordinate transform.

We cannot immediately insert $E_0(z)$ into the left hand side of Eq.~(\ref{eq:onAxisField}) and solve for $S'_0$ because each Bessel mode propagates with a field described by $J_m(k_rr)e^{i(k_zz+m\phi-\omega t)}$, which means it has a fast on-axis oscillation in $z$ of wavenumber $k_z$. Adding a fast carrier wave of wavenumber $k_{z0}$ results in a target on-axis field of
\begin{equation} \label{eq:targetOnAxisField}
    E(0, z, t) = E_0(z)e^{ik_{z0}(z)z}E_t(t-z/v)e^{-i\omega_0(t-z/v)},
\end{equation}
where $v$ is the group velocity of the pulse---the selection of which determines the pulse front tilt necessary at $z=0$---and we have allowed $k_{z0}(z)$ to be $z$ dependent. From Eqs.~(\ref{eq:krkz}) and (\ref{eq:besselBeam}), we can see that the width of the focus, parameterized by the radius to the first Bessel zero $R_b$, is related to the choice of $k_{z0}(z)$ by
\begin{equation} \label{eq:BesselWidth}
    R_b(z)\approx \frac{2.4048}{\sqrt{k^2-k_{z0}(z)^2}},
\end{equation}
which determines the amount of energy required in the initial laser pulse; a wider focus increases the energy required.

Combining Eq.~(\ref{eq:onAxisField}) with Eq.~(\ref{eq:targetOnAxisField}), we can solve for $S'_0(k_z, \omega)$ to get
\begin{equation} \label{eq:spatialSpectrum}
    S'_0(k_z, \omega)=\frac{1}{(2\pi)^2 k_z}\hat{E}_t(\omega_0-\omega)\int_{-\infty}^\infty dz\,E_0(z)e^{i[k_{z0}(z)+(\omega-\omega_0)/v]z}e^{-ik_zz},
\end{equation}
where a hat denotes the Fourier transform: $\hat{E}_t(\omega)=\int_{-\infty}^\infty E_t(t)e^{-i\omega t}\,dt$. Only modes with $k_z$ in the interval $\left<0, k\right>$ are propagating. This bandwidth limitation determines how faithfully $E_0(z)$ can be recreated by the optical system, as shown by Eq.~(\ref{eq:onAxisField}) \cite{ouadghiri2016arbitrary}. Note that, via the Fourier shift theorem, $k_{z0}$ shifts the Fourier transform of $E_0(z)$ towards $k$; as a result, larger $k_{z0}$ (larger spot size $R_b$) will reduce the bandwidth available.

For the special case where $k_{z0}(z)=k_{z0}$, Eq.~(\ref{eq:spatialSpectrum}) simplifies to
\begin{equation}
    S'_0(k_z, \omega)=\frac{1}{(2\pi)^2 k_z}\hat{E}_t(\omega_0-\omega)\hat{E}_0\left(k_z-k_{z0}-\frac{\omega-\omega_0}{v}\right).
\end{equation}
The Bessel mode corresponding to $k_{z0}$ is composed of a cone of rays with an angle of $\cos\theta=k_z/k$ relative to the optical axis. The velocity of the pulse in the direction of the optical axis, for a diffractive optic, is $v=ck_{z0}/k_0$, where $k_0=\omega_0 n_g/c$. For a reflective optic, or refractive optic with negligible variation in group delay, $v=ck_0/k_{z0}$.

\begin{figure}[htbp]
  \centering
  \includegraphics[width=3.37in]{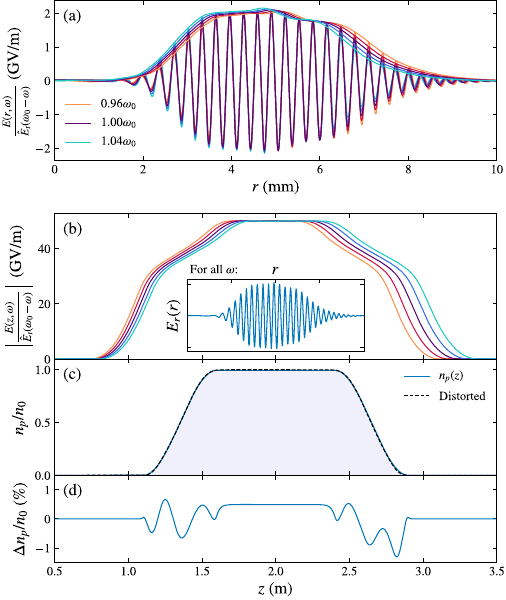}
  \caption{Plasma density error due to neglecting the chromatic dependence of the electric field at $z=0$. (a) The electric field on the $z=0$ plane required to produce the target electric field of Fig.~\ref{fig:IonizationFraction}(c) with a \qty{50}{fs} pulse, $R_b=\qty{120}{\micro\meter}$, and a velocity of $v=ck_{z0}/k_0$. The field is calculated at several frequencies using Eq.~(\ref{eq:E_r_omega}) to show the chromatic dependence. The envelope of the field is shown to help visualize the shift with frequency. (b) If all frequencies have the same field at $z=0$ (inset), then the on-axis field strength varies with frequency, distorting the pulse shape. (c) The corresponding plasma density resulting from the distorted pulse shape. (d) The error in the plasma density relative to $n_0$, a negligible error for PWFA applications. \label{fig:IgnoringChromatics}}
\end{figure}

The field on the $z=0$ plane is found by transforming back to $k_r$ with $S_0(k_r, \omega)=S'_0\big(\sqrt{k^2-k_r^2}, \omega\big)$ and setting $z=0$ in Eq.~(\ref{eq:besselBeam}). This results in
\begin{equation} \label{eq:targetField}
    E(r, 0, t) = \int_{-\infty}^\infty d\omega\int_0^\infty k_r\,dk_r\,S_0(k_r, \omega) J_0(k_rr)e^{-i\omega t}.
\end{equation}
Taking the Fourier transform gives the field as a function of frequency
\begin{equation} \label{eq:E_r_omega}
    E(r, \omega) = 2\pi\int_0^\infty k_r\,dk_r\,S_0(k_r, \omega) J_0(k_rr).
\end{equation}
It is difficult to produce an arbitrary field for all frequencies; however, as shown in Fig.~\ref{fig:IgnoringChromatics}(a), the field is similar for all $\omega$, allowing us to make a monochromatic approximation and evaluate Eq~(\ref{eq:E_r_omega}) at only $\omega_0$ to get the radial field that needs to be generated by the optical system: 
\begin{equation}\label{eq:targetEr}
    E_r(r) = \int_0^\infty k_r\,dk_r\,S_k(k_r) J_0(k_rr),
\end{equation}
where $S_k(k_r)=2\pi S'_0\left(\sqrt{k_0^2-k_r^2}, \omega_0\right)/\hat{E}_t(0)$. The lens design task has now been reduced to designing an optical setup that transforms the initial field $\mathcal{N}E_\perp$ into $E_r$. The consequences of the monochromatic approximation are shown in Fig.~\ref{fig:IgnoringChromatics}. Some chromaticity is introduced into the focal region, slightly distorting the pulse shape, but the resulting plasma density profile remains largely unaffected. The distortion increases as the distance between the lens and the plasma is increased. 

\begin{figure}[htbp]
  \centering
  \includegraphics[width=5.25in]{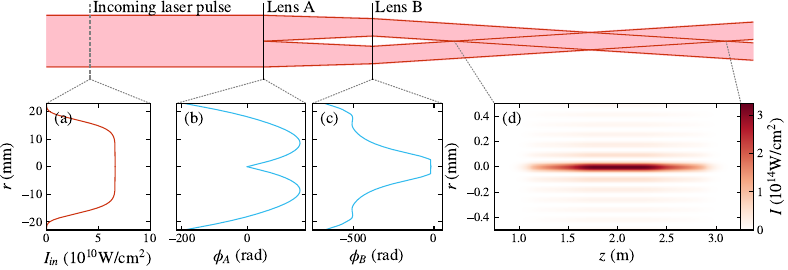}
  \caption{Tandem lens system for laser pulse shaping. Lens A shapes the incoming laser intensity profile (a) into the required intensity profile at Lens B, positioned at $z=0$. Lens B cancels out the residual phase imparted by Lens A and adds the necessary phase to produce the target on-axis electric field shown in Fig.~\ref{fig:IonizationFraction}(c), with a \qty{50}{fs} pulse, $v=ck_{z0}/k_0$, and $R_b=\qty{60}{\micro\meter}$. The phase profiles of Lenses A and B (b and c) are found using Eqs.~(\ref{eq:phaseLensA}) and (\ref{eq:phaseLensB}), respectively. The laser pulse is focused by the lens pair into a Bessel beam (d), achieving the required intensity profile along the optical axis. \label{fig:TandemLens}}
\end{figure}

To create the requisite field, we use the tandem lenses described in Ref.~\cite{honkanen1998tandem}. The tandem lens optical setup consists of two diffractive lenses, denoted Lens A and Lens B, separated by a distance $L$, as shown schematically in Fig.~\ref{fig:TandemLens}. Lens A reshapes the incoming laser intensity to the desired intensity at $z=0$. Lens B, placed at $z=0$, removes the phase added by Lens A and imparts the phase required to match the target field. The phase of Lens A is determined using a ray optics approach, which requires that the energy in the incoming pulse matches the energy in the target pulse:
\begin{equation} \label{eq:IinNorm}
    \mathcal{N}^2 = \frac{\int_{-\infty}^\infty rdr \left|E_r(r)\right|^2}{\int_{-\infty}^\infty rdr \left|E_\perp(r)\right|^2}.
\end{equation}

\begin{figure}[htbp]
  \centering
  \includegraphics[width=3.37in]{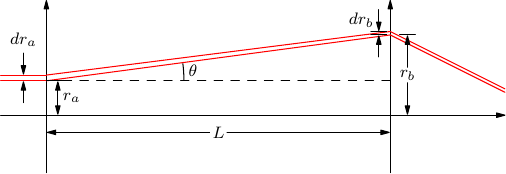}
  \caption{Energy conservation approach for finding the phase of Lens A. The energy incident on an annulus of radius $r_a$ and thickness $dr_a$ at Lens A propagates to an annulus of radius $r_b$ and thickness $dr_b$ at Lens B, following a ray that travels at an angle $\theta$. \label{fig:LensLayout}}
\end{figure}

The Lens A design problem is then to find the ray deflection angle $\theta_0(r)$ that maps the incoming laser intensity at Lens A, $I_{in}=\frac{1}{2}c\epsilon_0\left|\mathcal{N}E_\perp(r)\right|^2$, to the output laser intensity at Lens B, $I_{out}=\frac{1}{2}c\epsilon_0\left|E_r(r)\right|^2$. Each radius on the lens, $r_a$, can be mapped to a radius on the target plane, $r_b$, with the geometric energy conservation argument shown in Fig.~\ref{fig:LensLayout}. The energy incident on an annulus of thickness $dr_a$ on the lens plane corresponds to the energy incident on an annulus of thickness $dr_b$ on the output plane,
\begin{equation} \label{eq:energyRays}
    2\pi I_{in}(r_a)r_a\,dr_a = 2\pi I_{out}(r_b)r_b\,dr_b.
\end{equation}
Which can be solved numerically for $r_a(r_b)$ with
\begin{equation}
    r_{a,i}^2 = 2\Delta r_b\frac{I_{out}(r_{b,i-1})r_{b,i-1}}{I_{in}(r_{a,i-1})} + r_{a,i-1}^2,
\end{equation}
where $\Delta r_b=r_{b,i}-r_{b,i-1}$. Typically, the intensity profile on the second lens is annular with negligible intensity at $r_b=0$, so it is necessary to begin the integration at an $r_{b,0}$ where the intensity becomes appreciable. Each ray goes from $r_{a,i}$ to $r_{b,i}$, with the ray angle $\theta_0$ given by:
\begin{equation} \label{eq:rayAngle}
    \tan\theta_0 = \frac{r_b-r_a}{L}.
\end{equation}
By selecting $r_{b,0}$ and the maximum $r_b$ such that $I_{in}>0$ and $I_{out}>0$, $r_a(r_b)$ is single-valued and can be inverted to get $r_b(r_a)$.

The phase of the electric field is related to the ray angle at Lens A by $\frac{d\phi}{dr} = k_0\sin\theta_0$ \cite{sochacki1992nonparaxial}. Integrating this expression yields the phase $\phi_A(r)$ that Lens A must impart to the laser,
\begin{equation} \label{eq:phaseLensA}
    \phi_A(r) = k_0\int_0^r dr_a\,\frac{r_b(r_a)-r_a}{\sqrt{L^2+[r_b(r_a)-r_a]^2}},
\end{equation}
with the lens acting as a thin phase mask with transmittance function $t_A(r)=\exp[i\phi_A(r)]$. Figure~\ref{fig:TandemLens}(b) shows an example of the typical phase profile of Lens A.

The phase of the electric field incident on Lens B, $\phi_{in}$, can similarly be found by using the ray angle at Lens B,
\begin{equation}
    \phi_{in}(r) = k_0\int_0^r dr_b\,\frac{r_b-r_a(r_b)}{\sqrt{L^2+[r_b-r_a(r_b)]^2}}.
\end{equation}
Lens B must cancel out this phase and add the phase of the target field $\phi_r(r)=\mathrm{arg}[E_r(r)]$, resulting in a lens phase of
\begin{equation} \label{eq:phaseLensB}
    \phi_B(r) = \phi_r(r) - \phi_{in}(r).
\end{equation}
The incoming phase of the electric field at Lens B can alternatively be calculated by applying the phase of Lens A to the incoming field and propagating it to Lens B with a diffraction integral. Either approach yields the same phase; an example of $\phi_B(r)$ is shown in Figure~\ref{fig:TandemLens}(c).

Note that it is also possible to use the same mathematical approach to make a hollow tube of ionized plasma---of potential interest for positron acceleration \cite{gessner2016demonstration}---by choosing $m>0$ and solving for $S_m(k_r,\omega)$ with $r$ at the maximum of $J_m(k_{r0}r)$ when finding Eq.~(\ref{eq:onAxisField}). The rest of the calculation then proceeds in a similar manner with a $\phi$ dependent phase appearing in Eq.~(\ref{eq:targetEr}) and thus on Lens B. 

\section{\label{sec:rayOptics}Ray Optics Calculation of Lens Efficiency}

Lens A and Lens B can be produced using lithographic etching to create binary diffractive optics. Diffractive optics have the advantage that they can be made very thin, minimizing the nonlinear phase accumulated in the transmissive optics, and are inexpensive to produce in small quantities, but they suffer from reduced efficiency due to diffraction losses. Here, we use a ray optics approach to calculate the transmission efficiency of the lens pair.

A binary diffractive optic imparts the target phase $\phi$ (modulo $2\pi$) at the design wavelength $\lambda_0$ using a surface relief height $d$, discretized into $N=2^l$ levels \cite{swanson1989binary}:
\begin{equation}
    d = \frac{d_{max}}{N}\left \lfloor{\frac{N}{2\pi}(\phi\,\mathrm{mod}\,2\pi)}\right \rfloor,
\end{equation}
where $d_{max}=\lambda_0/[n_L(\lambda_0)-1]$ is the depth required for a $2\pi$ phase delay and $n_L(\lambda)$ is the index of refraction of the lens material. The optic can be produced using $l$ etching steps. 

Fabricated as diffractive optics, the optical system will have losses due to (i) Fresnel reflection off the front and back surfaces of each lens, given in the paraxial approximation by \cite{born2013principles}
\begin{equation}
    \eta_T(\lambda) \approx \frac{4n_L(\lambda)}{[n_L(\lambda)+1]^2}.
\end{equation}
(ii) Scattering to higher diffractive orders due to the discretization of the phase, (iii) improper surface relief height for wavelengths other than $\lambda_0$, and (iv) improper surface relief height for non-normal incidence on the diffractive surface of the second lens. The diffraction efficiency into the first order, which includes effects (ii)-(iv), is given by the scalar theory of Ref.~\cite{swanson1991binary}:
\begin{equation}
    \eta_D(\phi_0) = \mathrm{sinc}^2\left(\frac{\pi}{N}\right)\left[\frac{\sin[\pi(1-\phi_0)]}{N\sin[\pi(1-\phi_0)/N]}\right]^2,
\end{equation}
where $\phi_0$ is the phase advance, in waves, induced by a surface relief of $d_{max}$, given by
\begin{equation} \label{eq:phaseAdvance}
    \phi_0 =\frac{\lambda_0}{\lambda[n_L(\lambda_0)-1]}\left(\sqrt{n_L(\lambda)^2-\sin^2\theta}-\cos\theta\right),
\end{equation}
where $\theta$ is the angle of incidence on the surface. In the paraxial approximation, $\phi_0$ reduces to
\begin{equation} \label{eq:paraxialPhi}
    \phi_0 \approx \alpha-\frac{d_{max}}{2\lambda n_L(\lambda)}\theta^2,
\end{equation}
where $\alpha$ is the relative phase acquired for wavelengths other than $\lambda_0$ \cite{buralli1989optical}:
\begin{equation}
    \alpha=\frac{\lambda_0[n_L(\lambda)-1]}{\lambda[n_L(\lambda_0)-1]}.
\end{equation}

\begin{figure}[htbp]
  \centering
  \includegraphics[width=3.37in]{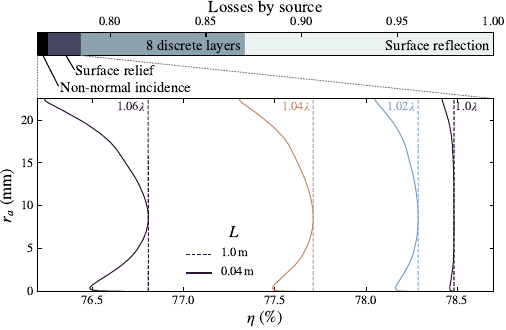}
  \caption{Efficiency breakdown of a binary optic tandem lens system using eight-layer fused silica optics for \qty{800}{nm}. The top bar shows the contribution from each loss source, with primary losses due to Fresnel reflections at the lens surfaces and phase discretization---both of which are improvable. The bottom plot shows the efficiency for off-wavelength components (different colors, due to improper surface relief height at those wavelengths) and rays with non-normal incidence. Curves are shown for two different lens separations. \label{fig:LensEfficiency}}
\end{figure}

A ray starting with normal incidence at Lens A with radial position $r_a$ will reach Lens B with angle and radius
\begin{equation} \label{eq:rayLensB}
\begin{split}
    \sin\theta &= \frac{\lambda}{\Delta_a} = -\frac{\lambda}{2\pi}\frac{d\phi_A}{dr}\\
    r_b &= r_a+L\tan\theta.
\end{split}
\end{equation}
where $\Delta$ is the grating period, related to the phase of the diffractive optic by
\begin{equation} \label{eq:gratingPeriod}
    \frac{1}{\Delta} = \frac{1}{2\pi}\abs{\frac{d\phi}{dr}}.
\end{equation}
The ray will lose energy due to reflections at all four interfaces, as well as diffraction on the two patterned surfaces. Utilizing Eqs.~(\ref{eq:paraxialPhi}), (\ref{eq:rayLensB}), and (\ref{eq:gratingPeriod}), the total efficiency is
\begin{equation}
    \eta(r_a, \lambda) = \eta_T(\lambda)^4\eta_D(\alpha)\eta_D\left(\alpha-\frac{d_{max}}{2n_L(\lambda)}\frac{\lambda}{\Delta_a^2}\right),
\end{equation}
where $\Delta_a$ is the grating period of the first lens. Figure~\ref{fig:LensEfficiency} shows the contribution to the efficiency of each source of loss. The largest contribution is the reflection losses, which can be mitigated by anti-reflective coatings, followed by the number of levels in the phase discretization. 

On the choice of L in the paraxial limit: The efficiency depends on $L$ through $d_{max}\lambda/(2n_L\Delta_a^2)$, which can be written by combining Eqs.~(\ref{eq:phaseLensA}) and (\ref{eq:gratingPeriod}) to give
\begin{equation} \label{eq:efficiencyL}
    \frac{d_{max}\lambda}{2n_L\Delta_a^2} = \frac{d_{max}\lambda}{2n_L(\lambda)\lambda_0^2}\frac{[r_b(r_a)-r_a]^2}{L^2+[r_b(r_a)-r_a]^2},
\end{equation}
with $r_b(r_a)$ referring to the ray with wavelength $\lambda_0$ found using Eq.~(\ref{eq:energyRays}). Note that $r_b$ is independent of $L$, which can be shown by combining Eqs.~(\ref{eq:phaseLensA}) and (\ref{eq:rayLensB}):
\begin{equation}
    r_b\approx r_a\left(1-\frac{\lambda}{\lambda_0}\right) + \frac{\lambda}{\lambda_0}r_{b0}(r_a).
\end{equation}
Efficiency is maximized when $d_{max}\lambda/(2n_L\Delta_a^2)$ is small compared to one, which results in the requirement $L\gg r_b(r_a)-r_a$, which is equivalent to the paraxial approximation. This requirement also guarantees that the size of the lens features is much larger than the laser wavelength, a requirement for the scalar theory of Sec.~\ref{sec:lensDesign} to be valid. As shown in Fig.~\ref{fig:LensEfficiency}, even for $L/[r_b(r_a)-r_a]\approx5$ the loss in efficiency is relatively small.

\section{\label{sec:simPerf}Simulated Focusing of an Ultrashort Pulse}

\begin{figure}[htbp]
  \centering
  \includegraphics[width=3.37in]{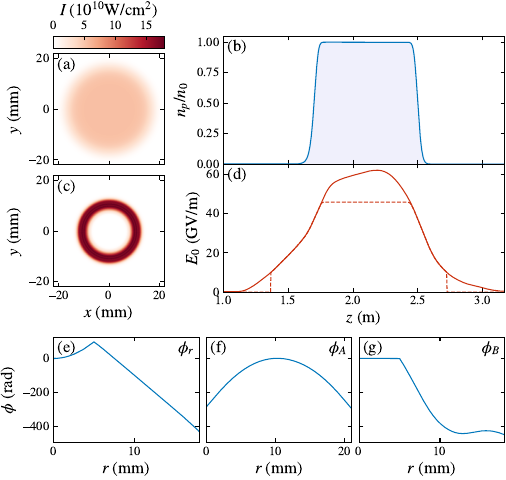}
  \caption{Design of the lenses used in the simulations and the experiment. (a) The input super-Gaussian laser profile. The input pulse duration is $\tau_p=\qty{50}{fs}$. (b) The target plasma density in $\mathrm{H_2}$. (c) The required intensity on the $z=0$ plane. (d) The field amplitude required along the optical axis to ionize the target plasma density, calculated by numerically inverting Eq.~(\ref{eq:ionizationFrac2}) (dashed). The target field was smoothed (solid) before calculating the lens phases. (e) The phase required on the $z=0$ plane, found using Eq.~(\ref{eq:targetEr}) with width $R_b=\qty{60}{\micro\meter}$. The phase of Lens A (f) and Lens B (g) was then calculated for $L=\qty{1}{m}$ using Eqs.~(\ref{eq:phaseLensA}) and (\ref{eq:phaseLensB}), respectively.  \label{fig:LO8LensDesign}}
\end{figure}

Diffractive optics are highly chromatic, resulting in phase and intensity errors for off-wavelength components after passing through the tandem lens system. The broad bandwidth of an ultrashort pulse will thus lead to distortion of both the on-axis fluence and pulse shape, causing an error in plasma density as compared to $n_p(z)$. To investigate these effects, we simulated the expected performance of the optical system using a cylindrically symmetric Fourier optics method; see App.~\ref{app:sim} for details \cite{guizar2004computation,oubrerie2022axiparabola}. The lens design used for both the simulations and the experiments is detailed in Fig.~\ref{fig:LO8LensDesign}. The lenses are designed to ionize a $\mathrm{H_2}$ gas, assuming that a single electron per molecule is ionized with an ionization energy of $\xi_i=\qty{15.4}{eV}$. This has been shown to be the dominant ionization pathway for \qty{800}{nm} laser pulses with $<\qty{55}{fs}$ pulse duration \cite{nie2022cross}.

The optical system performance was simulated using a pulse with a super-Gaussian transverse profile, given by $I_{in} = I_0 \exp\left[-2(r/w_0)^n\right]$, with $w_0=\qty{19}{mm}$, order $n=8$, and $I_0=\qty{5.43e10}{W/cm^2}$ found using Eq.~(\ref{eq:IinNorm}). The pulse shape was Gaussian with a FWHM duration of \qty{35}{fs}, to match the bandwidth of the experimental laser, for a total pulse energy of \qty{25}{mJ}. For all frequency components, $\phi_0=1$. As such, the simulation does not include any sources of loss at the optics. As discussed above, losses due to surface reflections and the discrete layers dominate; taking them into account, a \qty{32}{mJ} pulse is required to match the simulation. The simulation used 23 frequency components from \qty{744.8}{nm} to \qty{862.7}{nm} to resolve the temporal pulse structure. The pulse was propagated through three sections with different grid sizes: the \qty{1}{m} section from Lens A to Lens B in 100 longitudinal steps using a radial grid extending out to \qty{30}{mm}, the \qty{1.25}{m} section from Lens B to the start of the focal region in 125 steps with an \qty{18}{mm} extent, and through the focal region with 350 steps and a \qty{10}{mm} extent. All transverse grids had 2048 cells.

\begin{figure}[htbp]
  \centering
  \includegraphics[width=3.37in]{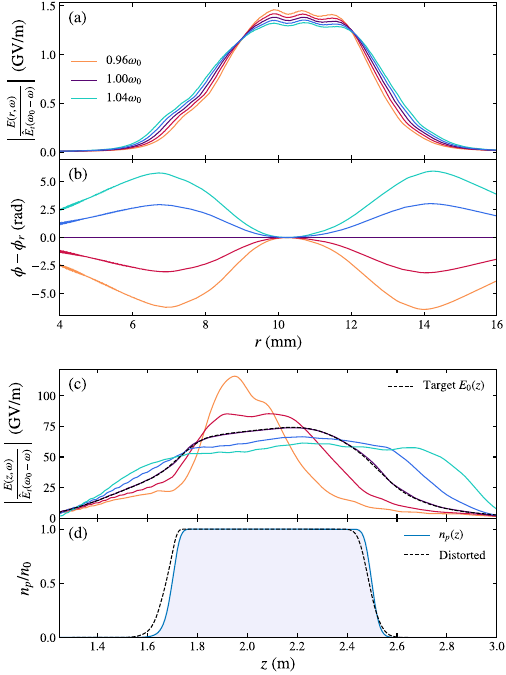}
  \caption{Simulated lens performance. (a, b) The field after Lens B for several frequency components. The variation is due to the diffractive nature of Lens A. (a) The magnitude of the field. (b) The difference between the phase after Lens B and the target phase. A uniform phase offset has been subtracted from each curve to align them at $r_0$. (c) The chromatic dependence of the on-axis field strength as a result of the phase error present after passing through the lenses. The electric field at the design wavelength closely matches the target electric field strength (dashed black). (d) The expected on-axis plasma density, calculated using the ADK expression, is only slightly distorted despite the significant chromatic aberrations in the tandem lenses. \label{fig:SimulatedPerformance}}
\end{figure}

The results of the simulation are shown in Fig.~\ref{fig:SimulatedPerformance}. From Eq.~(\ref{eq:rayLensB}) it can be seen that longer wavelengths will have a larger ray deflection after Lens A and thus form a narrower annulus at Lens B, as shown in Fig.~\ref{fig:SimulatedPerformance}(a). While the effect on the field amplitude is small, the phase of Lens B no longer perfectly cancels the residual phase from Lens A for frequencies other than $\omega_0$, as shown in Fig.~\ref{fig:SimulatedPerformance}(b). The error can be estimated by approximating the phase of Lens A as quadratic around the parallel ray with coefficient $\phi_2$ [the ray located at $\approx\qty{10}{mm}$ in Fig.~\ref{fig:LO8LensDesign} (f)]: $\phi_A=-\frac{1}{2}\phi_2(r-r_0)^2$, where $r_0$ is the radius of the parallel ray. The lowest order phase error around $r_0$ is approximately
\begin{equation}
    \phi_{in}(r, \lambda)-\phi_{in}(r, \lambda_0) \approx -\frac{k_0\phi_2(r-r_0)^2(\lambda-\lambda_0)}{1+L\phi_2/k_0},
\end{equation}
where a phase term constant in $r$ has been dropped. This quadratic phase term explains the variation in the on-axis field for different frequencies seen in Fig.~\ref{fig:SimulatedPerformance}(c). The phase error adds additional focusing to long wavelengths, which results in a shorter depth of focus, while shorter wavelengths experience the opposite effect. When viewed in the spectral domain, the aberration is significant, but the impact on the pulse duration and fluence is minimal. Figure~\ref{fig:SimulatedPerformance}(d) shows that the plasma density of the laser-ionized plasma suffers only minor distortions. The overall plasma length and ramp length are similar to the target but slightly shifted. 

\section{\label{sec:exp}Lens Characterization}

\begin{figure*}[htbp]
  \centering
  \includegraphics[width=5.25in]{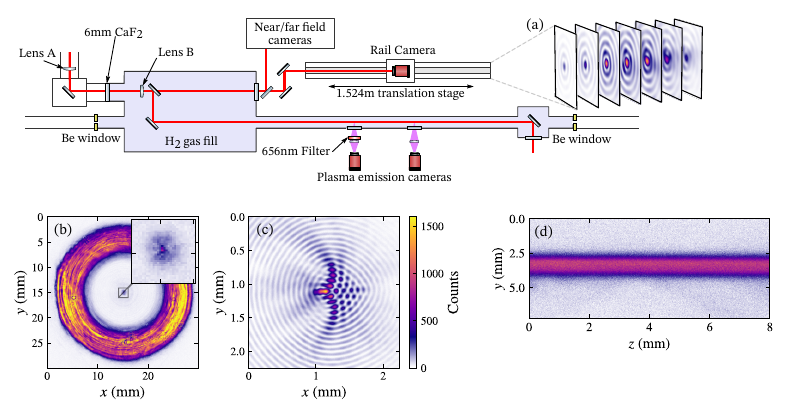}
  \caption{Experimental setup. A \qty{225}{mJ}, \qty{45}{fs} FWHM laser pulse was focused by a pair of tandem lens optics to produce a meter-scale Bessel focus. (a) The focal region was characterized by extracting the leakage light from a steering mirror and scanning a camera (Rail camera) longitudinally through the focal volume. The bulk of the laser reflected off a holed mirror to become co-linear with the electron beam, where it formed a plasma in the $\mathrm{H_2}$ gas-filled beamline. The gas was contained between two holed Be windows by a differential pumping system. (b) Lens A creates a small diffractive spot (inset), here imaged at the location of Lens B with the near field camera, used to align the lens to the laser. (c) Focal spot intensity measured \qty{1.95}{m} downstream of Lens B. Lens B is offset \qty{0.5}{mm} in the positive $x$-direction. The pattern is oriented horizontally, indicating the lenses are aligned vertically.(d) Light emission from the plasma verifying ionization of the $\mathrm{H_2}$ gas. \label{fig:09-ExperimentalSetup}}
\end{figure*}

Lens characterization took place at the FACET-II facility at SLAC National Accelerator Laboratory \cite{yakimenko2019facet}, demonstrating that the system can be integrated into the accelerator infrastructure necessary for a PWFA. The FACET-II experimental area is equipped with a \qty{10}{TW} Ti:sapphire laser system, located in a laser room above ground, that produces \qty{600}{mJ}, \qty{800}{nm} laser pulses at \qty{10}{Hz} \cite{green2014laser}. The pulses pass through a waveplate-polarizer attenuator before reflecting off a deformable mirror for wavefront correction. They then travel from the laser room to the pulse compressor, located in the accelerator housing, through a ~\qty{30}{m} long vacuum transport. A pair of image relay telescopes in the transport line, plus one in the laser room, image the output of the final amplifier to the input of the pulse compressor chamber. In the compressor, the pulse duration is compressed to \qty{45}{fs} FWHM with an output energy of \qty{225}{mJ} before being delivered to the experimental area. An overview of the experimental area is shown in Fig.~\ref{fig:09-ExperimentalSetup}.

Lens A of the tandem lens was located at the exit of the compressor chamber, and Lens B was located \qty{1}{m} downstream, in an experimental chamber on the accelerator beamline. Between the two lenses was a \qty{6}{mm} thick removable $\mathrm{CaF_2}$ window that isolated the beamline vacuum from the compressor vacuum. After Lens B, the laser reflected off two mirrors---the second of which had a \qty{4}{mm} hole for the electron beam to pass through---to become collinear with the electron beam axis. Leakage light through the first mirror was split between two diagnostic arms: a near field/far field camera pair, where the near field imaged the plane of Lens B, and a camera---called the Rail Camera---on a \qty{1.524}{m} translation stage that viewed the laser focus directly on the camera sensor [see Fig.~\ref{fig:09-ExperimentalSetup}(a)]. 

Lens A was mounted on a two-axis transverse mover, used to align the lens to the incoming laser by observing the position of the central diffractive spot from Lens A on the near field camera imaging Lens B, as shown in Fig.~\ref{fig:09-ExperimentalSetup}(b). Lens A and B were sensitive to relative transverse alignments on the order of \qty{10}{\micro\meter}. When the lenses were greatly misaligned, the pattern in the focal region formed a distinctive shape whose orientation corresponds to the direction of misalignment. To achieve alignment to Lens A, Lens B was mounted on a three axis mover. To align the vertical (horizontal) direction, the lens was offset horizontally (vertically) by \qty{500}{\micro\meter} and translated vertically (horizontally) to remove any tilt in the pattern, as shown in Fig.~\ref{fig:09-ExperimentalSetup}(c). Once aligned, the deformable mirror was used to optimize the focal spot as seen on the Rail Camera with the camera positioned at the downstream end of the focus. The three dimensional fluence profile in the focal region was then measured by scanning the Rail Camera along the length of the focus with the laser energy reduced to \qty{2.7}{mJ} using the attenuator in the laser room. 

\begin{figure}[htbp]
  \centering
  \includegraphics[width=3.37in]{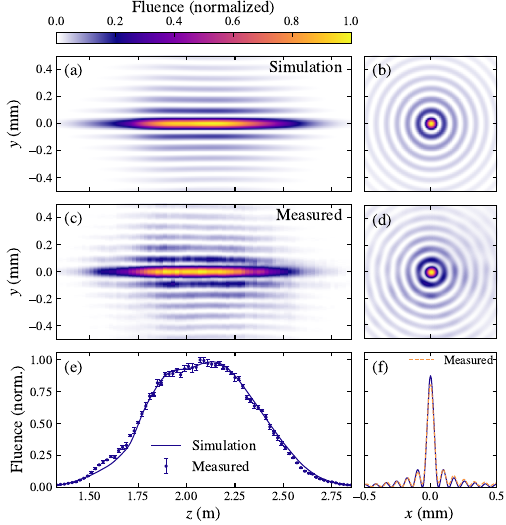}
  \caption{Measured lens performance. (a) Simulated fluence in the focal region of the tandem lenses and (b) on the $z=\qty{1.89}{m}$ plane. (c) Fluence in the focal region measured by scanning a camera along the length of the focus, and (d) Fluence measured with the camera \qty{1.89}{m} downstream of the second lens. The fluence is normalized to the maximum value in each image. (e) Comparison of the on-axis fluence from simulation and from fitting a squared $J_0$ Bessel function to each image, showing that the lens is producing the expected on-axis intensity profile. The simulation results have been scaled to match the measurements. (f) Measured Bessel spot size compared to the simulated spot size, created by taking line-outs in $x$ from (b) and (d). \label{fig:LensCharacterization}}
\end{figure}

Figure~\ref{fig:LensCharacterization} compares the measured fluence with the fluence predicted by the simulation of Sec.~\ref{sec:simPerf}. At each longitudinal position, ten images were taken, aligned, and then averaged to form Fig.~\ref{fig:LensCharacterization}(c, d). The measured on-axis fluence in Fig.~\ref{fig:LensCharacterization}(e) was found by fitting $I(r)=AJ_0(k_rr)^2$ to each two-dimensional image in order to separate the lens performance from the laser wavefront; the standard deviation of the amplitude from the ten images is shown. The agreement is excellent, demonstrating that the optics were shaping the laser as expected. The width of the Bessel spot was calculated from the fits using Eq.~(\ref{eq:BesselWidth}) and was within 3.2\% of the design value of \qty{60}{\micro\meter} along the full length of the focus.

A plasma was formed by filling the interaction point (IP) region with \qty{1.35}{torr} of hydrogen gas and increasing the laser energy to \qty{192}{mJ} to singly ionize the $\mathrm{H_2}$ molecules. The gas was contained in a \qty{4}{m} long region by the inner two apertures of a differential pumping system, which separated the linac vacuum from the IP without any beam-intercepting windows \cite{storey2024wakefield}. A continuous gas inflow replaces the outflow through the apertures, maintaining the pressure to within 1\% of the set-point. A camera---imaging the beam axis through a window on the side of the beamline---confirmed plasma formation by detecting light emission from the H-alpha line through a \qty{656.3}{nm}, \qty{10}{nm} FWHM line filter, as shown in Fig.~\ref{fig:09-ExperimentalSetup}(b). 

\section{\label{sec:conc}Conclusion}

We have experimentally demonstrated a novel optical setup that allows for precise control of the on-axis intensity profile of a high-power ultrafast laser system. The ability to shape the intensity opens the possibility of creating high repetition-rate, optically accessible plasma sources with tailored longitudinal density profiles. Such tailored density profiles are a necessary requirement for preserving the emittance of the particle beam in a plasma wakefield accelerator, allowing the acceleration of beams of sufficient quality for colliders and light sources. 

Our fully chromatic extension of the pulse shaping theory allows us to achieve excellent agreement with experiment and understand the chromatic effects of the optical system. The diffractive tandem lens demonstrated here introduces significant chromatic aberrations into the focused pulse compared to the slight wavelength dependence required for exact shaping. Simulations showed that control of the plasma profile at the level necessary for PWFA was still achievable despite this chromatic aberration. Better control could likely be achieved by using an iterative lens design approach to correct for errors in the plasma density. Another alternative would be to use reflective optics for both tandem lenses, with the added advantage of improving the efficiency of the optics.

The laser-ionized $\mathrm{H_2}$ plasma source has been successfully used in a PWFA to demonstrate record-high drive-to-wake energy transfer efficiency. The details of the experiment will appear in an upcoming publication.

\begin{backmatter}
\bmsection{Funding}
This work was supported by the U.S. Department of Energy under DOE Contract No. DE-AC02-76SF00515 and by the U.S. Department of Energy, Office of Science, Office of High Energy Physics under Award No. DE-SC0017906.
\bmsection{Data Availability Statement}
Data supporting the findings of this study are available from the corresponding author upon a reasonable request.
\end{backmatter}

\appendix

\section{\label{app:ADK}ADK Model}

The ADK model of Ref.~\cite{Ammosov1986} gives the ionization rate $w$ of an electron in terms of the magnitude of the electric field $E$. In SI units, the ADK formula is:

\begin{equation} \label{eq:usefulADK}
    w_{\!A\!D\!K}(E) = \abs{C_{n^*l^*}}^2 \frac{e}{\hbar} \frac{(2l+1)(l+\abs{m})!}{2^{\abs{m}}(\abs{m})!(l-\abs{m})!}\xi_i \left(\frac{2E_0}{E}\right)^{2n^*-\abs{m}-1}\exp\left(-\frac{2E_0}{3E}\right),
\end{equation}
where $l$ is the orbital angular momentum quantum number of the electron, $m$ is its projection, 
\begin{align}
    \abs{C_{n^*l^*}}^2 &= \frac{2^{2n^*}}{n^*\Gamma(n^*+l^*+1)(n^*-l^*)}, \\
    E_0 &= 2^{3/2}\frac{\sqrt{m_ee}}{\hbar}\xi_i^{3/2}, \\
    n^* &= \alpha c\sqrt{\frac{m_e}{2e}}\frac{Z}{\sqrt{\xi_i}},
\end{align}
with $l^*=n^*-1$, $\xi_i$ the ionization energy of the electron in electron volts, $Z$ the ionization level of the atom after removing the electron ($Z=1$ for the ionization of a neutral atom), $\alpha$ is the fine structure constant, and $\hbar$ is the reduced Planck constant.



\section{\label{app:sim}Fourier Optics Simulation}

The pulse propagation simulations are carried out using standard scalar Fourier optics techniques \cite{oubrerie2022axiparabola} in cylindrical coordinates using the discrete Hankel transform \cite{guizar2004computation}. The field on the $z=0$ plane is decomposed into Bessel modes, each with amplitude given by
\begin{equation}
    S(k_r, \omega) = \int_{-\infty}^\infty dt\int_0^\infty r\,dr\, E(r,t)J_0(k_rr)e^{-i\omega t}.
\end{equation}
The field at any point in space and time is then given by Eq.~(\ref{eq:besselBeam}).

The numerical algorithm proceeds as follows: The initial field is separable, allowing the radial and time component to be sampled onto separate grids, the radial with range $[0, R)$ and the temporal with range $[-T/2, T/2)$. The discrete Fourier transform is first applied to $E_t(t_i)$ to get $\hat{E}_t(\omega_i)$ and $N_f$ frequency components centered on $\omega_0$ are retained. Each Bessel mode has an amplitude $S_{\omega_i, k_{r,j}}=\hat{E}_t(\omega_i)H_{jk}E_r(r_k)$. The Hankel transform matrix $H$ is found by numerically inverting the non-symmetric inverse Hankel Transform matrix $H_{ij}^{-1}=J_0(r_ik_{r,j})$, where $k_{r,j}=\alpha_j/R$ and $\alpha_i$ are the roots of $J_0$ \cite{guizar2004computation}. The field is then propagated along $z$ with 
\begin{equation}
    S_{\omega_i, k_{r,j}}(z)=S_{\omega_i, k_{r,j}}(z_0)e^{ik_z(z-z_0)},
\end{equation}
with $k_z=\sqrt{\omega^2/c^2-k_r^2}$. The field in real space is found by applying the inverse Hankel transform $E_{r_j}(\omega_i)=T_{jk}S_{\omega_i, k_{r,k}}$, followed by the inverse Fourier transform in the temporal direction.

\bibliography{bibliography.bib}

\begin{thebibliography}{10}
\newcommand{\enquote}[1]{``#1''}

\bibitem{hogan2005multi}
M.~J. Hogan, C.~D. Barnes, C.~E. Clayton, \emph{et~al.}, \enquote{Multi-gev energy gain in a plasma-wakefield accelerator,} {\protect\JournalTitle{Physical Review Letters}} \textbf{95}, 054802 (2005).

\bibitem{blumenfeld2007energy}
I.~Blumenfeld, C.~E. Clayton, F.-J. Decker, \emph{et~al.}, \enquote{Energy doubling of 42 gev electrons in a metre-scale plasma wakefield accelerator,} {\protect\JournalTitle{Nature}} \textbf{445}, 741--744 (2007).

\bibitem{adli2013beam}
E.~Adli, J.-P. Delahaye, S.~J. Gessner, \emph{et~al.}, \enquote{A beam driven plasma-wakefield linear collider: from higgs factory to multi-tev,} {\protect\JournalTitle{arXiv preprint arXiv:1308.1145}}  (2013).

\bibitem{foster2023hybrid}
B.~Foster, R.~D’Arcy, and C.~A. Lindstr{\o}m, \enquote{A hybrid, asymmetric, linear higgs factory based on plasma-wakefield and radio-frequency acceleration,} {\protect\JournalTitle{New Journal of Physics}} \textbf{25}, 093037 (2023).

\bibitem{asai2024exploring}
S.~Asai, A.~Ballarino, T.~Bose, \emph{et~al.}, \enquote{Exploring the quantum universe: Pathways to innovation and discovery in particle physics,} {\protect\JournalTitle{arXiv preprint arXiv:2407.19176}}  (2024).

\bibitem{nakajima2008towards}
K.~Nakajima, \enquote{Towards a table-top free-electron laser,} {\protect\JournalTitle{Nature Physics}} \textbf{4}, 92--93 (2008).

\bibitem{wang2021free}
W.~Wang, K.~Feng, L.~Ke, \emph{et~al.}, \enquote{Free-electron lasing at 27 nanometres based on a laser wakefield accelerator,} {\protect\JournalTitle{Nature}} \textbf{595}, 516--520 (2021).

\bibitem{pompili2022free}
R.~Pompili, D.~Alesini, M.~Anania, \emph{et~al.}, \enquote{Free-electron lasing with compact beam-driven plasma wakefield accelerator,} {\protect\JournalTitle{Nature}} \textbf{605}, 659--662 (2022).

\bibitem{labat2023seeded}
M.~Labat, J.~C. Cabada{\u{g}}, A.~Ghaith, \emph{et~al.}, \enquote{Seeded free-electron laser driven by a compact laser plasma accelerator,} {\protect\JournalTitle{Nature Photonics}} \textbf{17}, 150--156 (2023).

\bibitem{muggli1999photo}
P.~Muggli, K.~Marsh, S.~Wang, \emph{et~al.}, \enquote{Photo-ionized lithium source for plasma accelerator applications,} {\protect\JournalTitle{IEEE Transactions on Plasma Science}} \textbf{27}, 791--799 (1999).

\bibitem{litos2014high}
M.~Litos, E.~Adli, W.~An, \emph{et~al.}, \enquote{High-efficiency acceleration of an electron beam in a plasma wakefield accelerator,} {\protect\JournalTitle{Nature}} \textbf{515}, 92--95 (2014).

\bibitem{litos20169}
M.~Litos, E.~Adli, J.~Allen, \emph{et~al.}, \enquote{9 gev energy gain in a beam-driven plasma wakefield accelerator,} {\protect\JournalTitle{Plasma Physics and Controlled Fusion}} \textbf{58}, 034017 (2016).

\bibitem{oz2007ionization}
E.~Oz, S.~Deng, T.~Katsouleas, \emph{et~al.}, \enquote{Ionization-induced electron trapping in ultrarelativistic plasma wakes,} {\protect\JournalTitle{Physical Review Letters}} \textbf{98}, 084801 (2007).

\bibitem{vafaei2016evidence}
N.~Vafaei-Najafabadi, W.~An, C.~Clayton, \emph{et~al.}, \enquote{Evidence for high-energy and low-emittance electron beams using ionization injection of charge in a plasma wakefield accelerator,} {\protect\JournalTitle{Plasma Physics and Controlled Fusion}} \textbf{58}, 034009 (2016).

\bibitem{lindstrom2021energy}
J.~Garland, G.~Boyle, P.~Gonzalez \emph{et~al.}, \enquote{Energy-spread preservation and high efficiency in a plasma-wakefield accelerator,} {\protect\JournalTitle{Physical Review Letters}} \textbf{126}, 014801 (2021).

\bibitem{lindstrom2024emittance}
C.~Lindstr{\o}m, J.~Beinortait{\.e}, J.~Bj{\"o}rklund~Svensson, \emph{et~al.}, \enquote{Emittance preservation in a plasma-wakefield accelerator,} {\protect\JournalTitle{Nature Communications}} \textbf{15}, 6097 (2024).

\bibitem{pena2024energy}
F.~Pe{\~n}a, C.~A. Lindstr{\o}m, J.~Beinortait{\.e}, \emph{et~al.}, \enquote{Energy depletion and re-acceleration of driver electrons in a plasma-wakefield accelerator,} {\protect\JournalTitle{Physical Review Research}} \textbf{6}, 043090 (2024).

\bibitem{deng2019generation}
A.~Deng, O.~Karger, T.~Heinemann, \emph{et~al.}, \enquote{Generation and acceleration of electron bunches from a plasma photocathode,} {\protect\JournalTitle{Nature Physics}} \textbf{15}, 1156--1160 (2019).

\bibitem{xu2016physics}
X.~Xu, J.~Hua, Y.~Wu, \emph{et~al.}, \enquote{Physics of phase space matching for staging plasma and traditional accelerator components using longitudinally tailored plasma profiles,} {\protect\JournalTitle{Physical Review Letters}} \textbf{116}, 124801 (2016).

\bibitem{litos2019beam}
M.~Litos, R.~Ariniello, C.~Doss, \emph{et~al.}, \enquote{Beam emittance preservation using gaussian density ramps in a beam-driven plasma wakefield accelerator,} {\protect\JournalTitle{Philosophical Transactions of the Royal Society A}} \textbf{377}, 20180181 (2019).

\bibitem{ariniello2019transverse}
R.~Ariniello, C.~Doss, K.~Hunt-Stone, \emph{et~al.}, \enquote{Transverse beam dynamics in a plasma density ramp,} {\protect\JournalTitle{Physical Review Accelerators and Beams}} \textbf{22}, 041304 (2019).

\bibitem{zhao2020emittance}
Y.~Zhao, W.~An, X.~Xu, \emph{et~al.}, \enquote{Emittance preservation through density ramp matching sections in a plasma wakefield accelerator,} {\protect\JournalTitle{Physical Review Accelerators and Beams}} \textbf{23}, 011302 (2020).

\bibitem{ariniello2022chromatic}
R.~Ariniello, C.~Doss, V.~Lee, \emph{et~al.}, \enquote{Chromatic transverse dynamics in a nonlinear plasma accelerator,} {\protect\JournalTitle{Physical Review Research}} \textbf{4}, 043120 (2022).

\bibitem{zgadzaj2020dissipation}
R.~Zgadzaj, T.~Silva, V.~Khudyakov, \emph{et~al.}, \enquote{Dissipation of electron-beam-driven plasma wakes,} {\protect\JournalTitle{Nature Communications}} \textbf{11}, 4753 (2020).

\bibitem{gilljohann2019direct}
M.~Gilljohann, A.~D{\"o}pp, J.~G{\"o}tzfried, \emph{et~al.}, \enquote{Direct observation of plasma waves and dynamics induced by laser-accelerated electron beams,} {\protect\JournalTitle{Physical Review X}} \textbf{9}, 011046 (2019).

\bibitem{lee2024temporal}
V.~Lee, R.~Ariniello, C.~Doss, \emph{et~al.}, \enquote{Temporal evolution of the light emitted by a thin, laser-ionized plasma source,} {\protect\JournalTitle{Physics of Plasmas}} \textbf{31} (2024).

\bibitem{vcivzmar2009tunable}
T.~{\v{C}}i{\v{z}}m{\'a}r and K.~Dholakia, \enquote{Tunable bessel light modes: engineering the axial propagation,} {\protect\JournalTitle{Optics Express}} \textbf{17}, 15558 (2009).

\bibitem{lu2022tunable}
Z.~Lu, Z.~Guo, M.~Fan, \emph{et~al.}, \enquote{Tunable bessel beam shaping for robust atmospheric optical communication,} {\protect\JournalTitle{Journal of Lightwave Technology}} \textbf{40}, 5097--5106 (2022).

\bibitem{fontaine2019attenuation}
Q.~Fontaine, H.~Hu, S.~Pigeon, \emph{et~al.}, \enquote{Attenuation-free non-diffracting bessel beams,} {\protect\JournalTitle{Optics Express}} \textbf{27}, 30067--30080 (2019).

\bibitem{yan2021non}
W.~Yan, Y.~Gao, Z.~Yuan, \emph{et~al.}, \enquote{Non-diffracting and self-accelerating bessel beams with on-demand tailored intensity profiles along arbitrary trajectories,} {\protect\JournalTitle{Optics Letters}} \textbf{46}, 1494--1497 (2021).

\bibitem{ouadghiri2016arbitrary}
I.~Ouadghiri-Idrissi, R.~Giust, L.~Froehly, \emph{et~al.}, \enquote{Arbitrary shaping of on-axis amplitude of femtosecond bessel beams with a single phase-only spatial light modulator,} {\protect\JournalTitle{Optics Express}} \textbf{24}, 11495 (2016).

\bibitem{honkanen1998tandem}
M.~Honkanen and J.~Turunen, \enquote{Tandem systems for efficient generation of uniform-axial-intensity bessel fields,} {\protect\JournalTitle{Optics Communications}} \textbf{154}, 368 (1998).

\bibitem{dharmavarapu2018diffractive}
R.~Dharmavarapu, S.~Bhattacharya, and S.~Juodkazis, \enquote{Diffractive optics for axial intensity shaping of bessel beams,} {\protect\JournalTitle{Journal of Optics}} \textbf{20}, 085606 (2018).

\bibitem{lai2017experimental}
Y.~H. Lai, J.~Xu, U.~B. Szafruga, \emph{et~al.}, \enquote{Experimental investigation of strong-field-ionization theories for laser fields from visible to midinfrared frequencies,} {\protect\JournalTitle{Physical Review A}} \textbf{96}, 063417 (2017).

\bibitem{Ammosov1986}
M.~Ammosov, N.~Delone, and V.~Krainov, \enquote{Tunnel ionization of complex atoms and of atomic ions in an alternating electromagnetic field,} {\protect\JournalTitle{Sov. Phys. JETP}} \textbf{64}, 1191 (1986).

\bibitem{perelomov1966ionization}
A.~Perelomov, V.~Popov, and M.~Terent’Ev, \enquote{Ionization of atoms in an alternating electric field,} {\protect\JournalTitle{Sov. Phys. JETP}} \textbf{23}, 924 (1966).

\bibitem{sochacki1992nonparaxial}
J.~Sochacki, A.~Ko{\l}odziejczyk, Z.~Jaroszewicz, and S.~Bara, \enquote{Nonparaxial design of generalized axicons,} {\protect\JournalTitle{Applied Optics}} \textbf{31}, 5326 (1992).

\bibitem{gessner2016demonstration}
S.~Gessner, E.~Adli, J.~M. Allen, \emph{et~al.}, \enquote{Demonstration of a positron beam-driven hollow channel plasma wakefield accelerator,} {\protect\JournalTitle{Nature Communications}} \textbf{7}, 11785 (2016).

\bibitem{swanson1989binary}
G.~J. Swanson \emph{et~al.}, \emph{Binary optics technology: the theory and design of multi-level diffractive optical elements}, vol. 854 (Massachusetts Institute of Technology, Lincoln Laboratory Cambridge, MA, USA, 1989).

\bibitem{born2013principles}
M.~Born and E.~Wolf, \emph{Principles of optics: electromagnetic theory of propagation, interference and diffraction of light} (Elsevier, 2013).

\bibitem{swanson1991binary}
G.~J. Swanson, \enquote{Binary optics technology: theoretical limits on the diffraction efficiency of multilevel diffractive optical elements,}  (1991).

\bibitem{buralli1989optical}
D.~A. Buralli, G.~M. Morris, and J.~R. Rogers, \enquote{Optical performance of holographic kinoforms,} {\protect\JournalTitle{Applied Optics}} \textbf{28}, 976--983 (1989).

\bibitem{guizar2004computation}
M.~Guizar-Sicairos and J.~C. Guti{\'e}rrez-Vega, \enquote{Computation of quasi-discrete hankel transforms of integer order for propagating optical wave fields,} {\protect\JournalTitle{JOSA A}} \textbf{21}, 53--58 (2004).

\bibitem{oubrerie2022axiparabola}
K.~Oubrerie, I.~A. Andriyash, R.~Lahaye, \emph{et~al.}, \enquote{Axiparabola: a new tool for high-intensity optics,} {\protect\JournalTitle{Journal of Optics}} \textbf{24}, 045503 (2022).

\bibitem{nie2022cross}
Z.~Nie, N.~Nambu, K.~A. Marsh, \emph{et~al.}, \enquote{Cross-polarized common-path temporal interferometry for high-sensitivity strong-field ionization measurements,} {\protect\JournalTitle{Optics Express}} \textbf{30}, 25696--25706 (2022).

\bibitem{yakimenko2019facet}
V.~Yakimenko, L.~Alsberg, E.~Bong, \emph{et~al.}, \enquote{Facet-ii facility for advanced accelerator experimental tests,} {\protect\JournalTitle{Physical Review Accelerators and Beams}} \textbf{22}, 101301 (2019).

\bibitem{green2014laser}
S.~Green, E.~Adli, C.~Clarke, \emph{et~al.}, \enquote{Laser ionized preformed plasma at facet,} {\protect\JournalTitle{Plasma Physics and Controlled Fusion}} \textbf{56}, 084011 (2014).

\bibitem{storey2024wakefield}
D.~Storey, C.~Zhang, P.~San Miguel~Claveria, \emph{et~al.}, \enquote{Wakefield generation in hydrogen and lithium plasmas at facet-ii: Diagnostics and first beam-plasma interaction results,} {\protect\JournalTitle{Physical Review Accelerators and Beams}} \textbf{27}, 051302 (2024).

\end{thebibliography}

\end{document}